\documentclass[12pt]{aastex} 
\usepackage{emulateapj5}

\newcommand{\be}{\begin{equation}}
\newcommand{\ee}{\end{equation}}

\begin{document}

\title{Spherical Accretion} 

\author{Re'em Sari$^{1,2}$ and Peter Goldreich$^{2,1}$}
\affil{$^1$Theoretical Astrophysics, Caltech 130-33, Pasadena, CA 91125}
\affil{$^2$Institute for Advanced Study, Einstein Drive, Princeton, NJ 08540}

\begin{abstract}
We compare different examples of spherical accretion onto a
gravitating mass. Limiting cases include the accretion of a
collisionally dominated fluid and the accretion of collisionless
particles. We derive expressions for the accretion rate and density
profile for semi-collisional accretion which bridges the gap between
these limiting cases. Particle crossing of the Hill sphere during the
formation of the outer planets is likely to have taken place in the
semi-collisional regime.
\end{abstract}

\section{Introduction}

We consider accretion onto a gravitating mass $M$ with an absorbing
boundary at radius $R$. The mass is embedded in a medium whose density
and velocity dispersion approach $\rho_{\infty}$ and $c_{\infty}$ as
$r\to \infty$. The attraction of the central mass is strongly felt
inside the gravitational radius at $r_g\equiv GM/c_{\infty}^2$. We
neglect the particles' self-gravity which is justified provided
$\rho_\infty r_g^3\ll M$.

We begin with a review of two examples, the accretion of a collision
dominated fluid in \S\ref{s:bondi} and the accretion of collisionless
particles in \S\ref{s:cless}.  A convenient measure of collisionality
is the optical depth across distance $r_g$ with $\rho=\rho_\infty$.
Given a scattering cross section per unit mass, or opacity\footnote{We
define opacity as a shorthand for collisional cross section per unit
mass. Its use does not imply that the transfer of radiation is
involved.}, $\kappa$,
\begin{equation}
\tau_g \equiv \kappa \rho_\infty r_g\, . 
\end{equation}
Semi-collisional accretion occurs for $\tau_g< 1$. 

We calculate new accretion rates and atmospheric density profiles for
steady-state, semi-collisional accretion of elastic and inelastic
particles in \S\ref{s:semi-elastic} and \S\ref{s:semi-inelastic},
respectively.  Astrophysical scenarios in which semi-collisional
accretion may apply are discussed in \S\ref{s:astro}.

\section{\label{s:bondi} Highly collisional (Bondi) accretion}

For $r>r_g$, $\rho(r)\sim \rho_{\infty}$, and for $r<r_g$, $v(r)\sim
(GM/r)^{1/2}$.  Therefore:
\begin{equation}
\dot M_B= 4 \pi\lambda r_g^2 \rho_{\infty} c_\infty\, .
\end{equation}
Note that $\dot M_B$ is independent of $R$. The dimensionless
coefficient $\lambda$ is of order unity. With a $\gamma$-law equation
of state, $\lambda=\exp(3/2)/4$ for $\gamma=1$ and $\lambda=1/4$ for
$\gamma=5/3$ \cite{Bondi52}.

For $r\ll r_g$, Bondi accretion is characterized by $v\propto
r^{-1/2}$ which implies $\rho\propto r^{-3/2}$ and hence $c\propto
r^{-3(\gamma-1)/4}$. Provided $\gamma<5/3$, the Mach number ${\cal
M}\equiv v/c\to \infty$ as $r/r_g\to 0$. Pressure gradients stymie
Bondi accretion for $\gamma>5/3$. The monatomic gas, $\gamma=5/3$, is
a special limiting case.

In Bondi accretion $\tau/\tau_g\approx (r_g/r)^{1/2}$ for $r< r_g$, so
collisionality increases inward. Thus $\tau_g\gg 1$ is a sufficient
condition for Bondi accretion provided $\gamma\leq 5/3$. 

\section{\label{s:cless}Collisionless Accretion}

A direct estimate of the accretion rate follows from the application of
the conservation laws for energy and angular momentum:
\begin{equation}
\label{Mdotcolless}
\dot M  \approx 2\pi
c_{\infty} \rho_\infty r_g R\approx (R/r_g){\dot M}_B\, .
\end{equation}
Contrary to the collisional Bondi case, the collisionless accretion rate
depends on $R$. 

At $r\ll r_g$, the density $\rho/\rho_\infty\sim (r_g/r)^{1/2}$, the
rms velocity dispersion is comparable to the free fall velocity, and
the net radial velocity is smaller by $\sim R/r$.

\section{\label{s:semi-elastic}Semi-Collisional Accretion of Elastic Particles}

Consider the steady accretion of particles that interact by elastic
scatterings under the condition that $\tau\equiv \kappa\rho r<1$ at
all $r_g>r>R$. A quasi-static atmosphere of bound particles with
$c^2\sim GM/r$ is present at $R<r<r_g$. It conducts constant
luminosity\footnote{Luminosity is defined as the rate at which energy
passes through a a certain radius. In this example the luminosity is
carried by conduction rather than by radiation.} $L\sim \tau r^2\rho
c^3$. The mass accretion rate $\dot M\sim r^2\rho v$ must also be
constant. ${\dot M}$ is determined by setting $v\sim c$ at $r=R$ since
particles that cross the inner boundary are absorbed. Thus $v/(\tau
c)\sim {\dot M}c^2/L\sim R/r$; although energy is transported outward
at speed $\sim \tau c$, the speed at which mass flows inward is a
factor $R/r\lesssim 1$ slower. It follows that
\begin{equation}
{{\dot M}\over {\dot M}_B}\sim {R\over r_g}\tau_g\, .
\end{equation}

Atmospheric profiles of density, $\rho(r)$, and entropy, $s(r)$,
depend on the opacity law which we parameterize as
$\kappa/\kappa_\infty=(c/c_\infty)^{-2\beta}$ with $\beta$ a
constant:
\begin{equation}
{\rho\over\rho_\infty}\sim \left(r_g\over r\right)^{3/4+\beta/2}\, ;
\end{equation}
\begin{equation}
s\propto \ln\left(c^3/\rho\right)\propto (\beta-3/2)\ln(r)\, .
\end{equation}
The entropy lost (for $\beta>3/2$) or gained (for $\beta<3/2$) by the
material as it moves toward smaller radius causes a minor variation
of the conductive luminosity:
\begin{equation}
{r\over L}{dL\over dr}\sim (\beta-3/2){R\over r} \ll 1\, ,
\end{equation}
where the strong inequality applies far outside the absorbing boundary,
i.e., for $r \gg R$.

\cite{BaW76} solved the important case of gravitational interactions
in connection with the accretion of stars by a black hole. Because the
cross section for strong scatterings is proportional to $c^{-4}$,
$\beta=2$. Thus $\rho\propto r^{-7/4}$. An earlier attempt by
\cite{Peebles72} yielded the incorrect result $\rho\propto r^{-9/4}$,
and herein lies an interesting lesson.  Peebles' assumed ${\dot M}\sim
\tau r^2\rho c$, whereas the correct result follows from setting
$L\sim \tau r^2\rho c^3$ \citep{Shapiro85}. In steady-state accretion,
both ${\dot M}$ and $L$ must be independent of $r$. However, as shown
above, the conduction of energy occurs at the maximum rate permitted
by relaxation whereas mass is transported more slowly.

\section{\label{s:semi-inelastic}Semi-Collisional Accretion of Inelastic 
Particles}

\subsection{accretion rate}

Collisions of unbound particles inside $r_g$ occur at relative
velocities of order $c_\infty(r_g/r)^{1/2}$.  Provided they dissipate
a significant fraction of the center of mass kinetic energy, they
produce bound particles. Comparable yields of bound particles come
from collisions between two unbound particles and from collisions
between an unbound and a bound particle. In either case, the addition
of bound particles comes mainly from collisions that occur near
$r_g$.\footnote{The statements in these last two sentences can be
verified after the density profile inside $r_g$ is determined.}
Subsequent collisions cause the bound particles to accrete. The total
accretion rate is bounded by the sum of the collisional and
collisionless accretion rates
\begin{equation}
\dot M\sim 2\pi c_{\infty} \rho_\infty r_g\left(\tau_g r_g+
R\right)\, .
\end{equation}
For $\tau_g> R/r_g$, collisions dominate the accretion rate and
\begin{equation}
{\dot M}\sim \tau_g{\dot M_B}
\label{dotM}
\end{equation}

\subsection{semi-collisional atmosphere}
\label{ss:atmos}

Next we turn our attention to the density profile in semi-collisional
atmospheres. Collisions involving an unbound particle serve as a
source for bound particles at a rate $\dot M\sim 2\pi\tau_g r_g^2
\rho_\infty c_{\infty}$. In steady state the density of bound
particles around $r_g$ is of the same order of magnitude as
$\rho_\infty$. It is set by a balance between the rate at which bound
particles are produced and the rate at which their mutual collisions
cause them to drift inward.

The mass accretion rate is independent of radius for $r \ll
r_g$. Together with knowledge of the average radial velocity, $v$,
this allows us to determine the density profile for constant
$\kappa$. Where $\tau\sim \kappa\rho r \lesssim 1$, $v/c_\infty \sim
\tau (r_g/r)^{1/2}$, but $v/c_\infty\sim (r_g/r)^{1/2}$ where
$\tau\gtrsim 1$. Thus
\begin{equation}
\label{rhonew}
\rho\approx \rho_\infty (r_g/r)^{5/4}\, ,
\end{equation}
if $\tau\lesssim 1$ and
\begin{equation}
\rho\approx \rho_\infty \tau_g(r_g/r)^{3/2}\,
\end{equation}
if $\tau\gtrsim 1$. For $R/r_g<\tau_g^4$, there is a transition
between the low and high optical depth regimes at $r/r_g\approx
\tau_g^4$.

We can now verify that collisions in the vicinity of $r_g$ dominate
the rate in which unbound particles become bound.  Even for the
steeper Bondi density profile, the rate of capture of particles
increases with $r$ for $r<r_g$;
\begin{equation}
\rho_{\rm un}(r)\rho_{\rm b}(r)c\kappa r^3=
\dot M_B \tau_g (r/r_g)^{1/2}\, ,
\end{equation}
where the subscripts $b$ and $ub$ refer to bound and unbound
particles, respectively. This approaches the semicollisional accretion
rate given by equation (\ref{dotM}) as $r\to r_g$.  Thus collisions
between an unbound and a bound particle make a contribution to the
accretion rate that is comparable to that made by collisions between
two unbound particles.\footnote{The presence of an atmosphere of bound
particles is a consequence of semi-collisional accretion, but it is
not the reason that the accretion rate is enhanced above the
collisionless value.}

For $\tau_g<R/r_g$ the semi-collisional atmosphere solution still
applies. However, its contribution to the accretion rate is
negligible. The formation timescale of the atmosphere is $\rho_\infty
r_g^3/ (\tau_g \dot M_B)$ whereas the accretion timescale is $M/(\dot
M_B R/r_g)$. Thus an atmosphere has sufficient time to form provided
\begin{equation}
\tau_g > {M_g \over M} \left. R \over r_g \right.
\end{equation}
where $M_g\approx \rho_\infty r_g^3$ is the mass of surrounding fluid
contained within the gravitational radius $r_g$.

Our results for the accretion rates as well as the bound particle atmosphere
are summarized in Table 1.
\begin{center}
\begin{table*}[ht!]
\begin{center}
\begin{tabular}{|c|c|c|c|c|c|}
\hline
                                          &   Bondi Accretion    & \multicolumn{2}{c|}{Partially collisional accretion}  &  \multicolumn{2}{c|}{Collisionless accretion} \\ \hline \hline
                                          &                 &    &  &  &            \\
         optical depth                    &   $\tau_g>1$    &  $1>\tau_g>\left( R\over r_g\right)^{1/4}$ & $\left(R\over r_g \right)^{1/4}>\tau_g>\left. R\over r_g\right.$ &  ${ R\over r_g} >\tau_g>\left. M_g\over M\right.\left. R\over r_g\right.$ & $\tau_g<\left.M_g\over M\right.\left. R\over r_g\right.$\\ 
                                          &                 &    &  &  &            \\ \hline
                                          &                 &     \multicolumn{2}{c|}{}  &   \multicolumn{2}{c|}{}            \\ 
         accretion rate                    &   $\dot M_B=4\pi \rho_\infty r_g^2 c_\infty$   & \multicolumn{2}{c|}{$\dot M_B \tau_g$}  & \multicolumn{2}{c|}{$\dot M_b R/r_g$} \\ 
                                          &                 &     \multicolumn{2}{c|}{}  &   \multicolumn{2}{c|}{}            \\ \hline
                                          &                 &    & \multicolumn{2}{c|}{}  &               \\ 
                                          &                 &   $\rho_\infty(r/r_g)^{-5/4}$ &  \multicolumn{2}{c|}{}  &            \\
         atmosphere         & $\rho_\infty(r/r_g)^{-3/2}$  &  & \multicolumn{2}{c|}{$\rho_\infty(r/r_g)^{-5/4}$}  & $\rho_\infty(r/r_g)^{-1/2}$            \\  
        &  & $\tau_g\rho_\infty(r/r_g)^{-3/2}$  & \multicolumn{2}{c|}{}  &            \\  
                                          &                 &    & \multicolumn{2}{c|}{}  &               \\  \hline \hline

\end{tabular}
\par
\label{t:summary}
\caption{ Summary of our results on semi-collisional accretion of inelastic particles.}
\end{center}
\end{table*}
\end{center}

\section{\label{s:astro}Semi-Collisional Planetesimal Accretion}

Consider the accretion of planetesimals by a protoplanet.  Our
idealized planetesimals are identical, indestructible, inelastic
spheres with radii $s$ and density $\rho_s$ so $\kappa \sim
3/s\rho_s$. Their velocity dispersion, $c_\infty$, is set by a balance
between excitation due to viscous stirring by protoplanets and damping
by mutual collisions. The notation in this section follows
that in \cite{GLS04}.

Suppose the planetesimal velocity dispersion is set at the boundary
between shear and dispersion dominated limits. Then the thickness of
the planetesimal disk is comparable to the planet's Hill radius,
$R_H$, which in turn is comparable to the gravitational radius,
$r_g$.\footnote{ $R_H\sim R/\alpha$, where $\alpha$ is approximately
the angular size of the sun as seen from the protoplanet's orbit.}
Denoting the surface mass density of planetesimals by $\sigma_\infty$,
we have $\tau_g \sim \tau_{\rm disk} \sim \sigma_\infty/s \rho_s$,
where $\tau_{\rm disk}$ is the vertical optical depth of the
planetesimal disk. Semi-collisional accretion applies for
$\alpha\lesssim \tau_g\lesssim 1$. All treatments of planet formation
of which we are aware are based on collisionless accretion. However,
this limit is appropriate only if the planetesimals are large
enough. For the fast growth of planets, which requires small
planetesimals, semi-collisional accretion may be the appropriate
regime. For example, a surface density of $\sigma_\infty \approx
1\,{\rm g/cm^2}$ is commonly adopted for the protoplanetary disk
around 30AU, where $\alpha \sim 10^{-4}$. With these values, the size
range for semi-collisional accretion is $1\,{\rm cm}<s<100\,{\rm m}$.

The collision rate inside the Hill sphere (cf. eq. [\ref{dotM}])
exceeds the collisionless accretion rate by $\sim\alpha^{-1}\tau_{\rm
disk}$. Each collision produces one or two bound particles. Spherical
symmetry is likely to be a poor approximation inside the Hill sphere.
The secular component of the Sun's tidal potential has a minimum in
the protoplanet's orbit plane. Collisions among bound particles damp
their motions perpendicular to this plane leading to the formation of
an accretion disk. It is unclear what fraction of these particles will
ultimately be accreted by the growing protoplanet.

\section{Summary}
The semi-collisional accretion of inelastic particles proceeds as
follows. Unbound particles that collide within the gravitational
radius, $r_g$, become bound and are ultimately accreted. This sets the
accretion rate.  The central mass is surrounded by a quasi-static
atmosphere of bound particles. Its density profile is determined by
taking ${\dot M}$ independent of $r$ and setting the mean inward
radial velocity equal to $\tau c$.  Where $\tau<1$ inside $r_g$, there
is an additional population of unbound particles, but its contribution
to the total density is minor. A bound atmosphere exists even at such
low $\tau_g$ that collisionless accretion dominates.  For extremely
low $\tau_g$, a steady state bound atmosphere cannot form on the
timescale in which the central body grows by collisionless
accretion. These results are summarized in Table 1.

Accretion rates and density profiles for the semi-collisional
accretion of inelastic particles differ from those for the
semi-collisionl accretion of elastic particles, such as
gravitationally interacting particles \citep{BaW76}. In the elastic
case, the majority of the bound particles are never accreted; they are
only temporarily captured. Ejections are a consequence of the outward
conduction of gravitational energy released mainly near the bottom of
the atmosphere.

Mutual collisions alter the nature of the accretion of small
planetesimals by protoplanets.  The formation of an accretion disk is
a likely outcome. If its density exceeds that of the unbound
particles, it would affect the formation and evolution of binaries in
scenarios such as that of \cite{GLS02}.

\acknowledgements We thank S. Shapiro for a helpful discussion. RS is
an Alfred P. Sloan Fellow and a Packard Fellow. This research was
supported in part by NASA Grant: NAG5-12037


\end{document}